\documentclass[letterpaper, twocolumn, 12pt]{article}

\usepackage{fullpage} 
\usepackage{parskip} 
\usepackage{tikz} 
\usepackage{amsmath}
\usepackage[colorlinks=true]{hyperref}
\usepackage{graphicx}
\usepackage{subfigure}
\usepackage{caption}
\usepackage{authblk}
\usepackage{setspace}

\usepackage{titlesec}
\usepackage{wrapfig}
\usepackage[utf8]{inputenc}
\usepackage[style=numeric, sorting=none]{biblatex}
\usepackage[symbol]{footmisc}

\titleformat{\section}
  {\normalfont\bfseries}{\thesection}{1em}{}
\titlespacing*{\section}{0pt}{3.25ex plus 1ex minus .2ex}{0ex plus .2ex}

\title{Lunar Opportunities for SETI}
\author{}
\date{}

\begin{document}

\begin{titlepage}
   \begin{center}
       \vspace*{1cm}

        \huge
        \textbf{Lunar Opportunities for SETI}
        
        

        
       \vspace{1.5cm}
        
        \normalsize
        Eric J. Michaud\footnote{Corresponding author. Phone: +1 (925) 596-8844. Email: eric.michaud99@gmail.com}$^{,1}$, Andrew P. V. Siemion$^{1, 2, 3}$, Jamie Drew$^{4}$, S. Pete Worden$^{4}$
        
        \vspace{0.3cm}
        
        \normalsize
        {\itshape 
            $^1$University of California Berkeley, Berkeley, CA 94720 \\
            $^2$SETI Institute, Mountain View, CA 94043 \\
            $^3$University of Malta, Institute of Space Sciences and Astronomy \\
            $^4$The Breakthrough Initiatives, NASA
            Research Park, Bld. 18, Moffett Field, CA, 94035
        }
        
    \vspace{0.9cm}
        \includegraphics[width=0.5\textwidth]{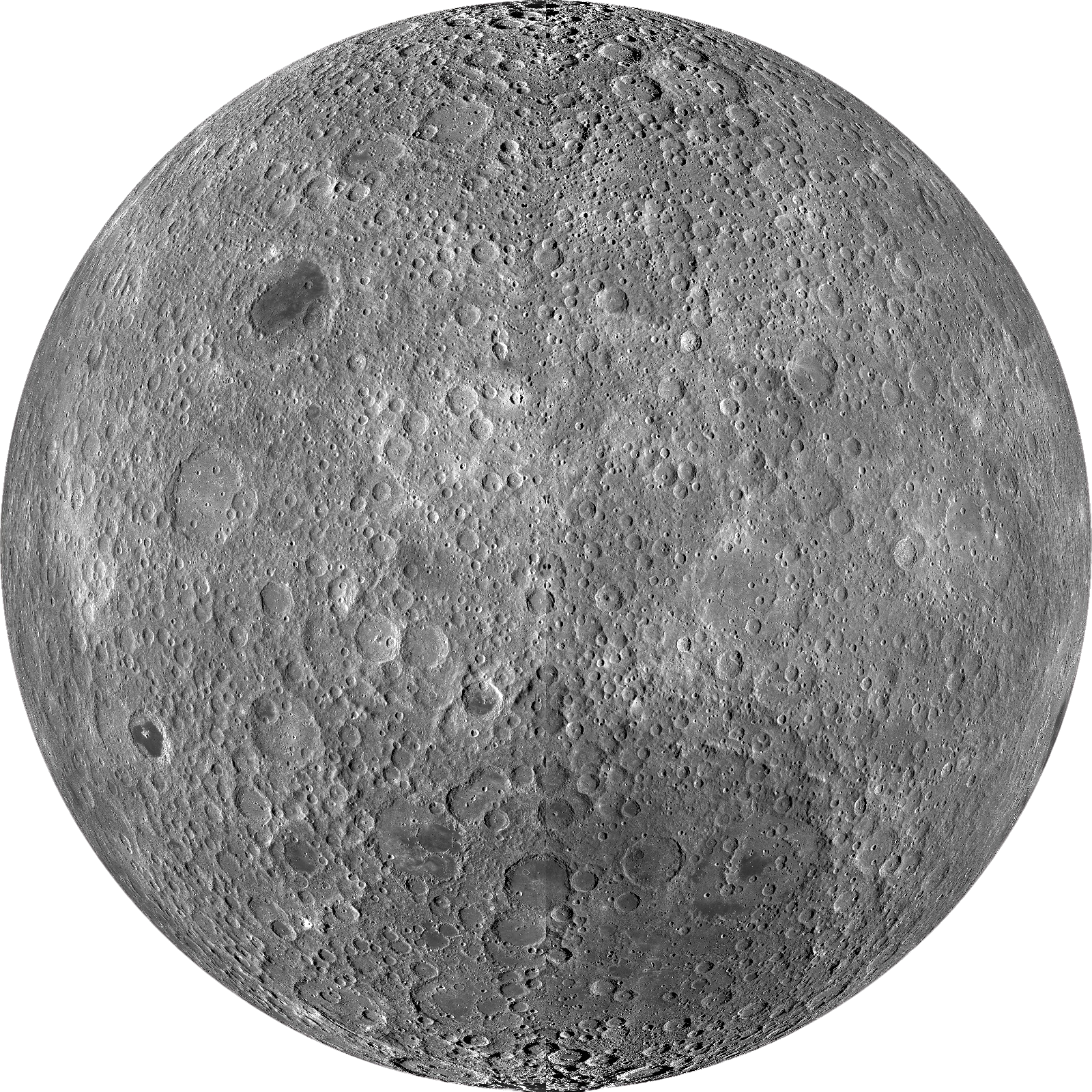}
        
    \vspace{0.9cm}
            
       A white paper \\ National Academy of Sciences \\ Planetary Science and Astrobiology Decadal Survey 2023-2032
       
    \vfill

            
   \end{center}
\end{titlepage}


\twocolumn[
  \begin{@twocolumnfalse}
    \begin{abstract}
      \bigbreak
      A radio telescope placed in lunar orbit, or on the surface of the Moon’s farside, could be of great value to the Search for Extraterrestrial Intelligence (SETI). The advantage of such a telescope is that it would be shielded by the body of the Moon from terrestrial sources of radio frequency interference (RFI). While RFI can be identified and ignored by other fields of radio astronomy, the possible spectral similarity between human and alien-generated radio emission makes the abundance of artificial radio emission on and around the Earth a significant complicating factor for SETI. A Moon-based telescope would avoid this challenge. In this paper, we review existing literature on Moon-based radio astronomy, discuss the benefits of lunar SETI, contrast possible surface- and orbit-based telescope designs, and argue that such initiatives are scientifically feasible, both technically and financially, within the next decade.

 \bigbreak
    \end{abstract}
  \end{@twocolumnfalse}
]

\section{Introduction}

Since the 1960s, many have recognized the unique opportunities for radio astronomy presented by the Moon \cite{gorgolewski, burnsconference, zarkaplanetary, lazio, falcke}. In 1986, Jack Burns et al. declared that “the Moon is very possibly the best location within the inner solar system from which to perform front-line astronomical research” \cite{burnsconference}. The Moon’s characteristics (its lack of an atmosphere, low seismic activity, long nights, etc.) make it attractive for a variety of astronomical projects. Notably, it would be a revolutionary platform for low frequency cosmology, which cannot be conducted from the surface of the Earth due to the shielding effects of our ionosphere \cite{lazio, daremission}. The search for extraterrestrial intelligence could be similarly transformed by a lunar radio telescope \cite{maccone2004seti, heidmann1996seti}. The primary advantage for SETI is that the body of the Moon provides an excellent shield against terrestrial radio frequency interference \cite{takahashithesis, pluchino, alexander}. 

\begin{figure}[!htb]
    \center{\includegraphics[width=0.37\textwidth]
    {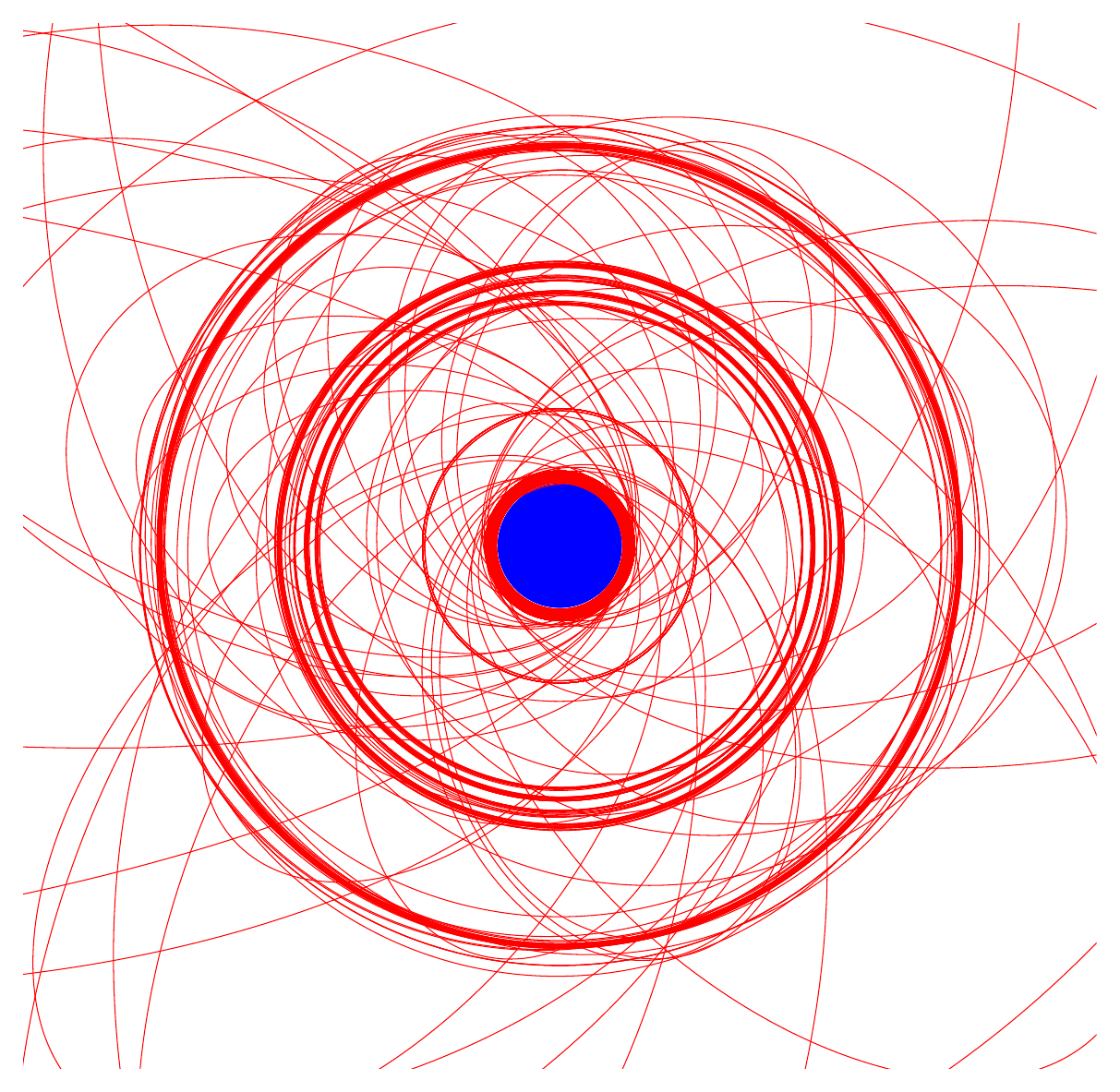}}
    \caption{\label{fig:orbits} Orbits of over 2000 active satellites around the Earth (ignoring inclination), generated from the UCS Satellite Database \cite{satellitedatabase}}.
\end{figure}

Searches for signs of alien technology, most notably for narrowband radio signals, are complicated by the abundance of human technology around the Earth. When human-produced radio emission is detected in high volume in SETI observations, it becomes challenging to attribute any particular signal to extraterrestrial intelligence. While SETI astronomers have developed observing strategies and specialized software for approaching this challenge \cite{2019arXiv190607750P}, it is worth exploring a more radical solution -- conducting observations from an area of space with minimal exposure to radio pollution.

\begin{figure*}[ht]
    \center{
        \includegraphics[height=3.5in]
        {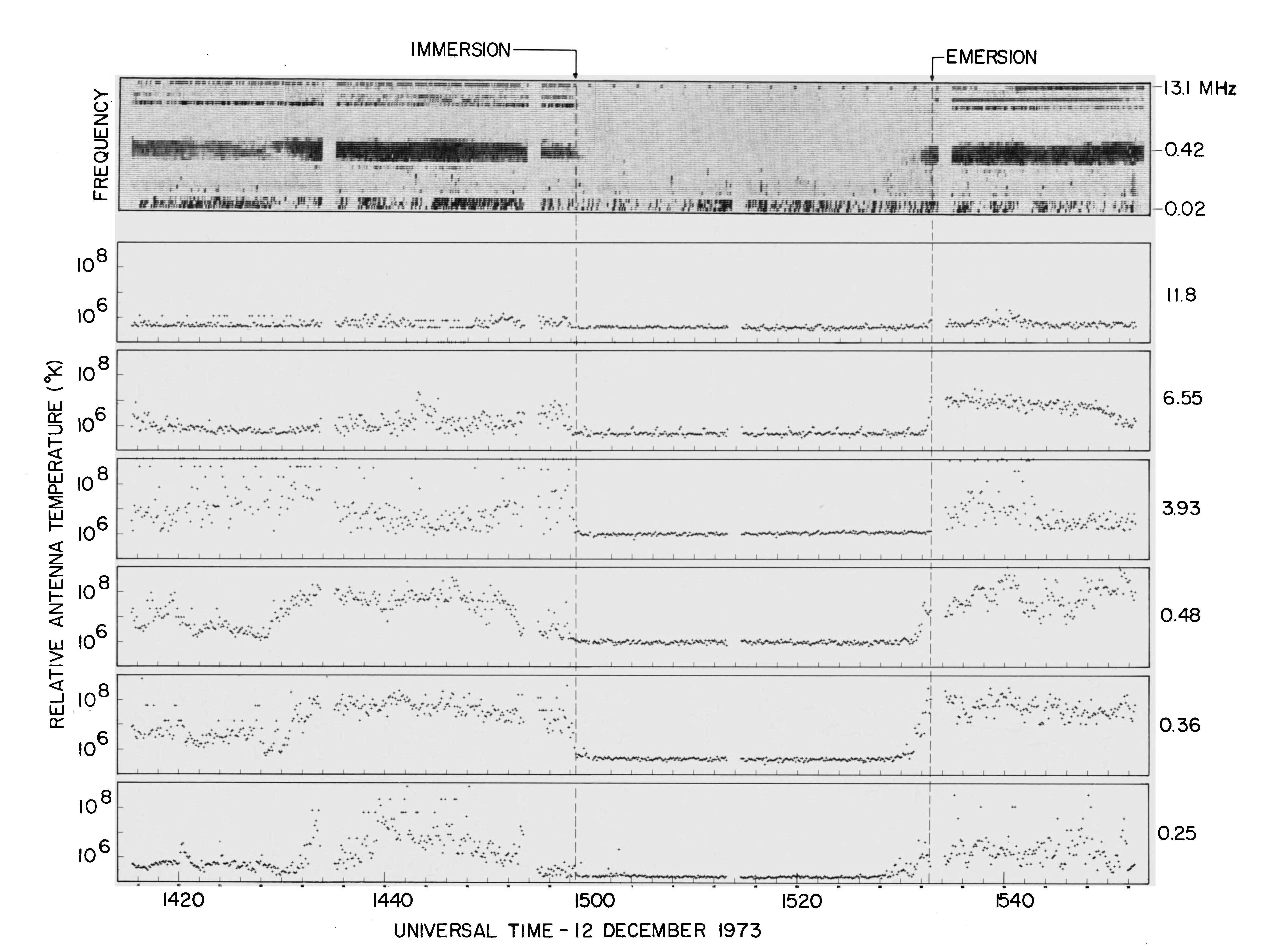}}
    \caption{\label{fig:alexander} Terrestrial radio interference as observed by the RAE-B satellite from lunar orbit. Note the clear drop in the power of terrestrial radio emission as the satellite passes behind the Moon. This shielding from terrestrial RFI would be ideal for SETI observations. Taken from \cite{alexander}
}
\end{figure*}

According to Zarka et al. “the farside of the Moon is during the Lunar night the most radio-quiet place of our local Universe” \cite{zarkaplanetary}. This assertion is supported both by computer simulations of radio-wave diffraction around the Moon \cite{takahashithesis, pluchino}, and by radio observations conducted in lunar orbit \cite{alexander}. 

In 1968, NASA’s RAE-A (Radio Astronomy Explorer) observed that from its Medium-Earth orbit, “radio emissions from the Earth – both natural and man-made – were very common and often very intense” \cite{alexander}. This finding motivated the placement of the subsequent RAE-B spacecraft into lunar orbit (at an altitude of 1100 km), where terrestrial noise would be blocked for some fraction of its orbit by the body of the Moon \cite{falcke}. RAE-B did indeed observe “impressive occultations” \cite{alexander} of such noise, as illustrated in Figure \ref{fig:alexander}. Terrestrial radio noise was attenuated by 1-3 orders of magnitude as the spacecraft passed behind the Moon.  

Computer simulations tell a similar story, and indicate that a radio telescope positioned on the surface of the lunar farside would be even more strongly protected from terrestrial RFI than a telescope in lunar orbit. In one such simulation, Yuki Takahashi \cite{takahashithesis} found that radio waves at frequencies as low as 50 kHz would likely be attenuated by at least 10 orders of magnitude on the opposite side. In another study, Pluchino et al. \cite{pluchino} estimated that near the crater Daedalus (almost exactly opposite the Earth), the power of geostationary satellite interference at 100 MHz and 100 GHz would be attenuated by 7 and 10 orders of magnitude respectively. 

\begin{figure}[!htb]
    \center{\includegraphics[width=0.34\textwidth]
    {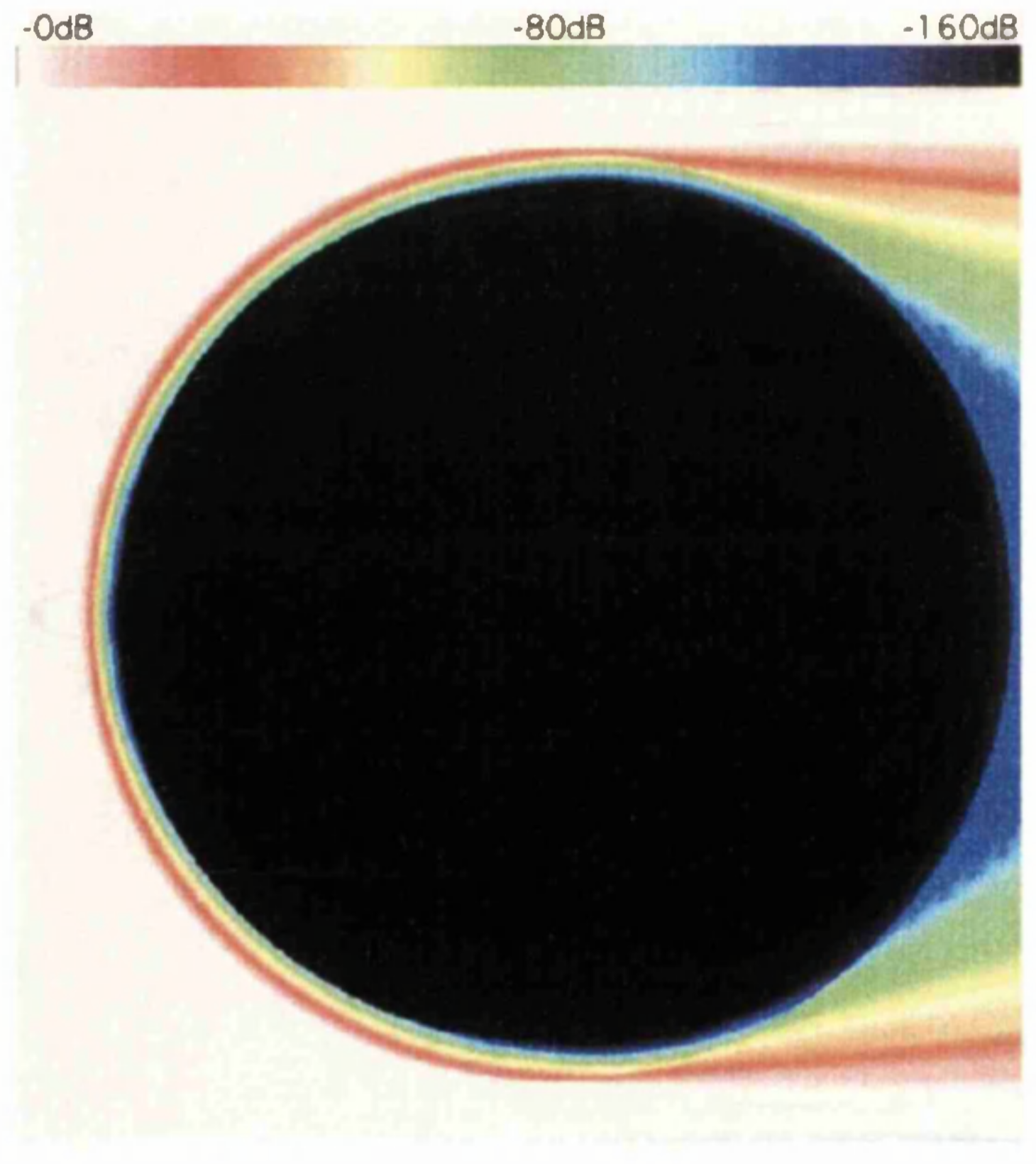}}
    \caption{\label{fig:takahashi} Simulated diffraction of a 60 kHz wave, incident from the left, around the body of the Moon. Over the farside, we see a predicted 10 order of magnitude attenuation of the signal power. Taken from \cite{takahashithesis}
}
\end{figure}

Together, these studies indicate that the radio environment around the Moon’s farside is radically different, and favorable in comparison with, the Earth’s surface or orbit for the purposes of SETI. Shielded from terrestrial interference, a Moon-based telescope would receive RFI only from satellites and rovers located on the Moon or beyond it. The number of such radio-producing devices is far lower than around the Earth, and even if human activity around the Moon were to increase substantially within the next decade, we should still expect that the number of such devices to be dramatically lower than the number of satellites in Earth orbit. (see Figure \ref{fig:orbits}). Nevertheless, it would be wise to protect the lunar farside as a radio quiet zone to the extent possible. Proposals like Claudio Maccone's \textit{Protected Antipode Circle} should be seriously considered by the international community \cite{pluchino, macconepac}.

\begin{figure*}
\centering     
\subfigure[RFI from the UHF Satcom]{\label{fig:a}\includegraphics[width=3in]{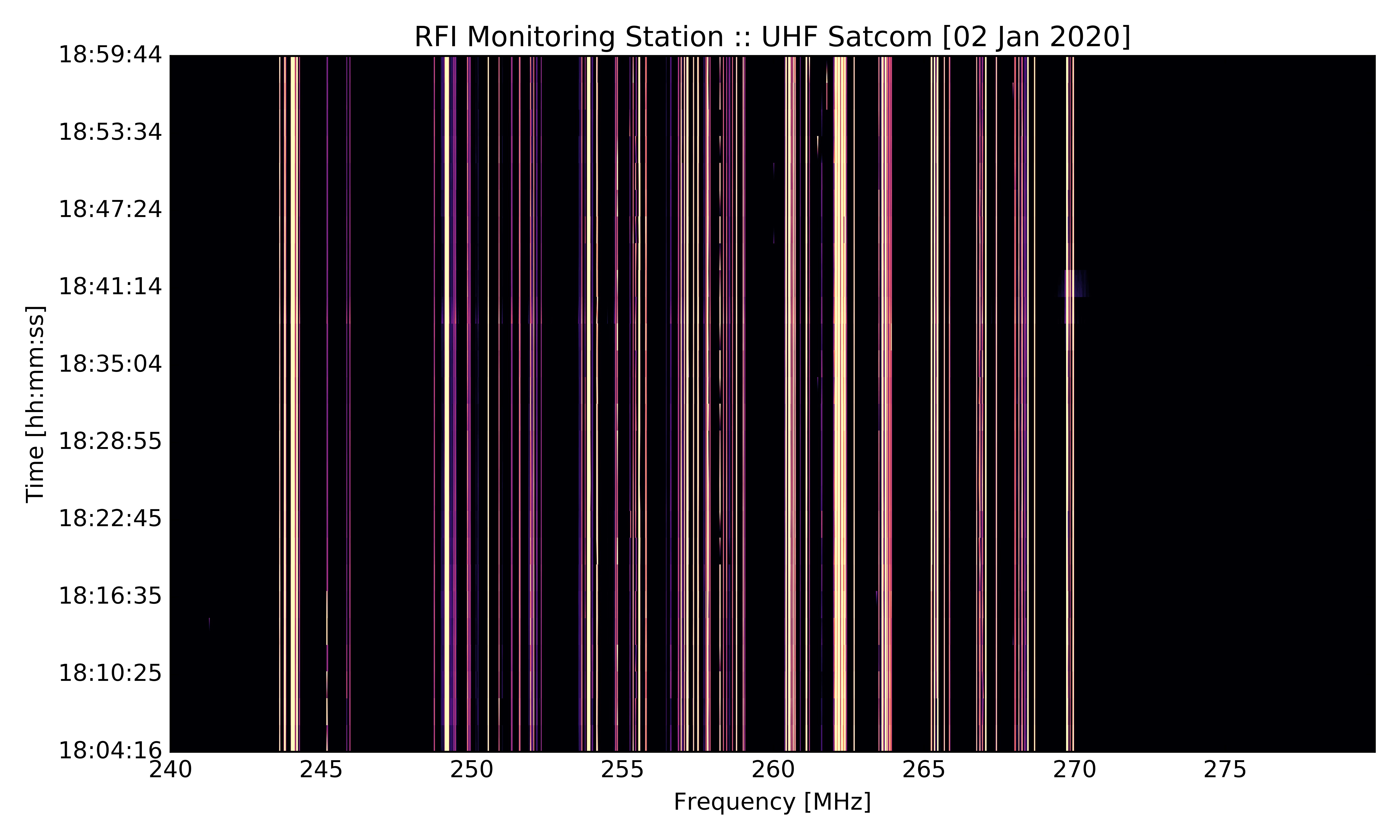}}
\subfigure[RFI from Iridium Satellites]{\label{fig:b}\includegraphics[width=3in]{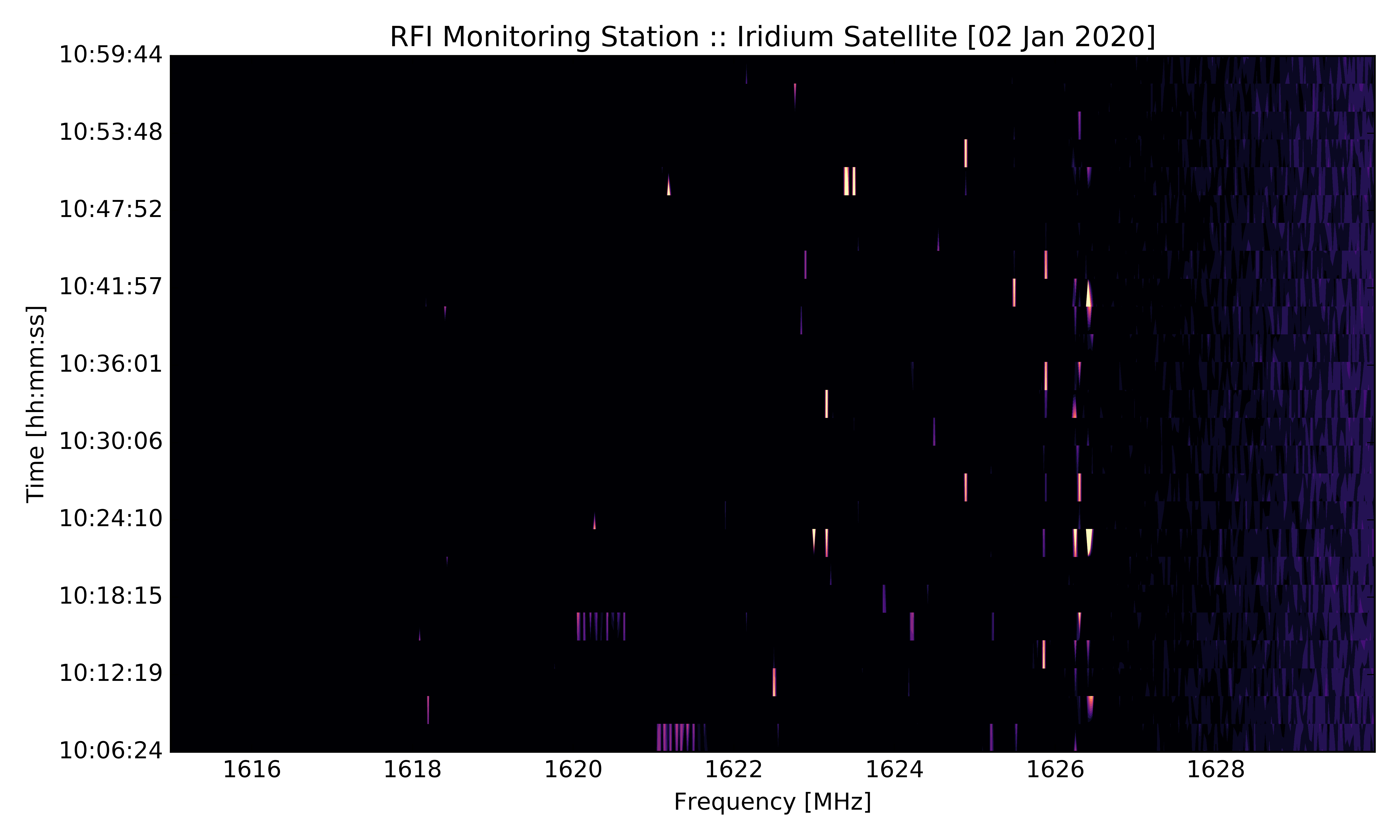}}
\caption{RFI observed from the MeerKAT telescope site in South Africa. The MeerKAT instrument will perform an unprecedented SETI survey of one million stars as part of the Breakthrough Listen Initiative. The presence of RFI at the telescope will increase the complexity of the data analysis associated with the campaign. A lunar observatory would not be exposed to this interference.}
\label{fig:meerkatrfi}
\end{figure*}


\section{Surface Mission Considerations}

Placing a radio telescope on the surface of the lunar farside would minimize its exposure to terrestrial RFI. Radio noise could be especially mitigated if a crater were selected as a landing site. The craters Saha \cite{heidmann, heidmann14}, Tsiolkovsky \cite{takahashithesis, burnstalk}, Malapert \cite{takahashithesis, takahashi15}, and Daedalus \cite{pluchino, macconepac} have been chosen in past lunar radio-astronomy proposals. Crater walls could block out interference originating from the Earth-Moon L4 and L5 points, as well as from lunar orbiters. The surface could also provide a stable platform for the construction of a larger dish, perhaps exploiting the curved geometry of a crater itself \cite{bandyopadhyay}. A set of smaller devices could also be positioned across the surface, forming an interferometer similar to LOFAR. The surface would also allow for longer continuous observations than an orbiter. To minimize the radio noise from the Sun, observations should be conducted during the $\sim$14 day lunar night \cite{zarkaplanetary, lazio}.

One possible constraint on a surface-based mission is that a large battery pack would likely be needed to power observations during the lunar night. Current solar-powered instruments on the farside, such as China’s Chang'e 4 lander and Yutu 2 rover, shut down during the night to conserve power. However, a radio telescope would be operational primarily during the night, and hence constraints on lander mass, and therefore on battery capacity, may limit the active observation time of the telescope to a fraction of the night. 

Communication would also be a challenge for a surface-based radio telescope. The obvious RFI benefit of never being in line-of-sight with Earth also makes it impossible to communicate without using a relay satellite. Currently, the only such satellite is China’s Queqiao, which orbits the Earth-Moon L2 point. In the next decade other options may become available for relaying communications.



\begin{figure*}
\centering     
\subfigure[]{\label{fig:c}\includegraphics[width=6in]{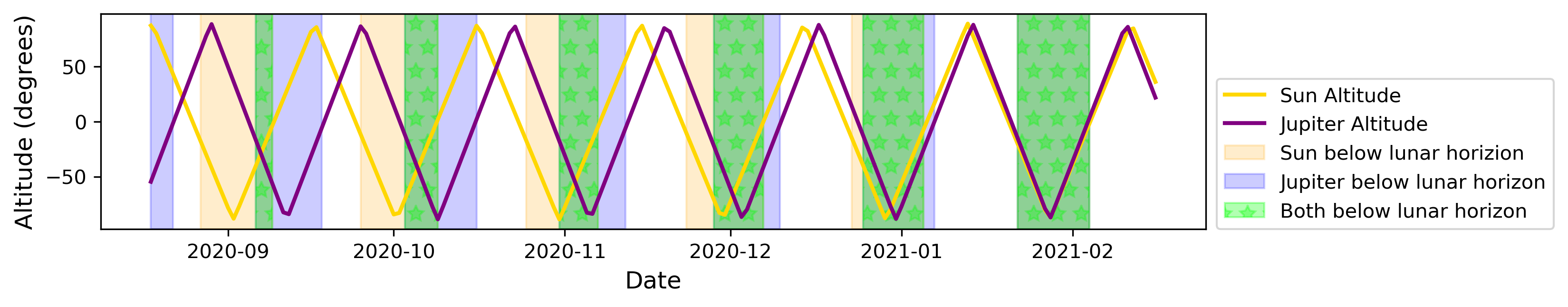}}
\subfigure[]{\label{fig:d}\includegraphics[width=6in]{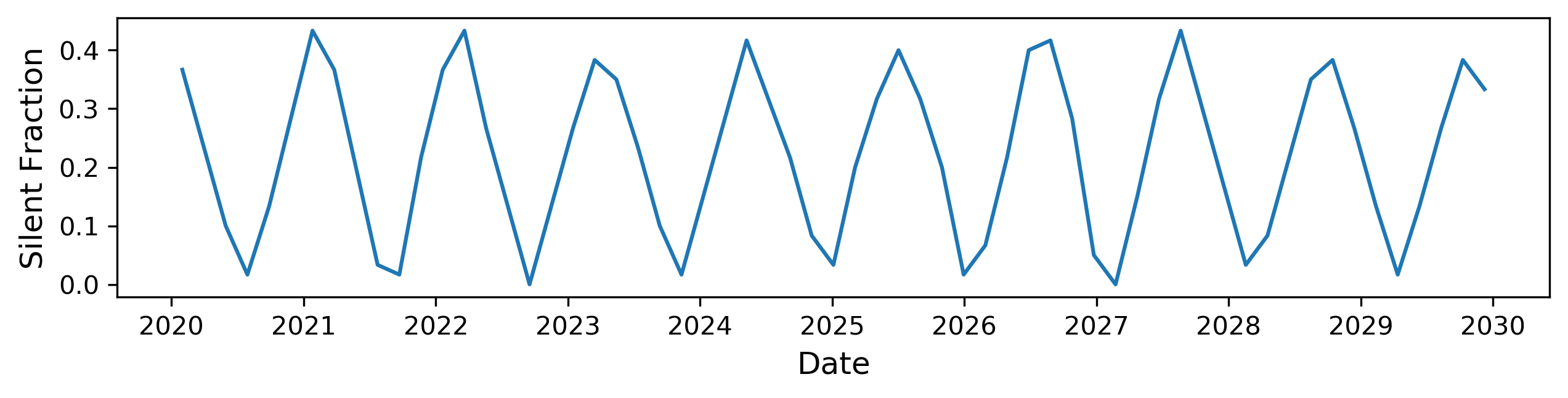}}
\caption{Jupiter being an emitter of low-frequency radio waves, some observations may be best conducted when both the Sun and Jupiter are obscured \cite{zarkaplanetary}. (a) displays altitude over time of the Sun and Jupiter for a telescope located at the lunar farside (0º N, 180º W). Green denotes periods where both the Sun and Jupiter are below the horizon. The length of this window varies throughout the year, as the relative positions of Jupiter and the Sun change from the perspective of the Earth-system. (b) depicts the fraction of time during which both bodies are obscured, averaged over 60-day windows, over the next ten years. Some periods of the year will be more favorable for observing than others. } 
\end{figure*}

\section{Orbital Mission Considerations}

A telescope in lunar orbit would not face the communications and power challenges of a surface-based one. Observations could be conducted while the orbiter is above the farside, and communications performed while over the near side. Orbiters are also less expensive to design and launch than landers - landing not only requires additional $\Delta v$ to complete, but also increases mission complexity and risk, as was recently shown in 2019 by the Israeli 'Beresheet' and Indian 'Vikram' lunar landing failures. An orbiter may also be able to support a larger antenna than a lander of equivalent mass, since the weightlessness of orbit obviates the need for certain structural elements that would be necessary at the surface under lunar gravity. There is already precedent for large radio telescopes in space. The RAE-2 orbiter had an impressive antenna length of 229 meters \cite{alexander}. Documents leaked by Edward Snowden revealed the details of high-altitude SIGINT satellites launched by U.S. intelligence services, a few of which feature a 20-30 meter unfurlable reflector dish \cite{sigint}.

There are some disadvantages to an orbiter-based SETI mission. RFI may might not be attenuated as much as on the farside surface. Also, due to the Moon's gravitational lumpiness, most lunar orbits are inherently unstable. For instance, the PFS-2 subsatellite (deployed during Apollo 16), lasted only 35 days before crashing into the surface \cite{bizarreorbits}. Fortunately, there exist several “frozen orbits” which could enable several years of observations from lunar orbit \cite{dono, folta}. Further work should be done on determining the best orbital parameters of such a mission.


\section{Additional Considerations}

When selecting between mission concepts, several additional questions should be considered. A major priority should be obtaining a more precise characterization of the lunar RFI environment. Particularly, how has China's Chang'e 4 mission affected the RFI environment of the farside? How could future lunar missions, like NASA's proposed Gateway space station, interfere with SETI observations? Also, what interference can we expect from artificial satellites and robotic landers elsewhere in the solar system? Furthermore, what might the hardware look like on a lunar radio telescope? 

\section{Feasibility and Timing}

Whether performed from the lunar surface or from orbit, Moon-based radio astronomy offers unique advantages for SETI. Critically, recent trends conspire to make such a mission not only increasingly feasible, but also increasingly necessary. The reduction in satellite launch costs \protect\cite{costreduction} and the popularization of smaller satellite buses is leading to an ever greater number of satellites being put in Earth orbit. SpaceX's StarLink constellation alone may contribute tens of thousands of new satellites to the already RFI-dense swarm around the Earth. This will further complicate Earth-surface-based SETI observation campaigns. However, the same economic and technological forces which are enabling this ramping up of satellite launches also make a lunar SETI mission more feasible. Small organizations now routinely place relatively inexpensive satellites into orbit. HawkEye 360, a small company based out of Virginia, has managed to design, build, and launch three satellites for the purpose of detecting and precisely locating radio sources on the surface of the Earth. These missions and others form a rough blueprint for, and signal the increasing feasibility of, sending a small instrument dedicated to SETI to the Moon. Such a mission would enable a detailed survey of the lunar RFI environment, and act as a proof of concept for more sophisticated missions in the future. A lunar SETI mission would mark the beginning of a new era in the history of SETI, where an increasing human presence in space is accompanied by an expanding ability to discover extraterrestrial life other than our own. 

\section{Acknowledgements}

We thank Claire Webb, David MacMahon, Steve Croft, Howard Isaacson, Julia DeMarines and especially Daniel Czech for their suggestions and feedback. Thanks to Braam Otto for contributing the waterfall plots for Figure \ref{fig:meerkatrfi}. This work was supported as part of the Breakthrough Listen Initiative, sponsored by the Breakthrough Prize Foundation.

{\footnotesize
\printbibliography
}

\end{document}